\documentclass[12pt]{article}
\pdfoutput=1

\usepackage{amssymb}
\usepackage{amsmath}
\usepackage{varwidth,xcolor}
\usepackage{graphicx}
\usepackage{subfig}
\usepackage{caption}
\usepackage{float}
\DeclareGraphicsExtensions{.pdf,.jpeg,.png,.eps,.jpg}

\begin{document}

\title{ Thermal conductance of suspended nanoribbons: interplay between strain and interatomic potential nonlinearity }

\author{ Roberto Barreto$^{1,2}$, M. Florencia Carusela $^{1,2,*}$,\\Alejandro G. Monastra$^{1,2}$}

\maketitle

\begin{center}
$^1$ Instituto de Ciencias, Universidad Nacional de Gral. Sarmiento, \\J. M. Guti\'errez 1150, (1613) Los Polvorines, Argentina\\
$^{2}$ Consejo Nacional de Investigaciones Cient\'\i ficas y T\'ecnicas, \\ Godoy Cruz 2290, (1425) Buenos Aires, Argentina
\end{center}


$^{*}$Email: flor@ungs.edu.ar

\renewcommand{\thesection}{\Roman{section}}	

\begin{abstract} 

We investigate the role that nonlinearity in the interatomic potential has on the thermal conductance of a suspended nanoribbon when it is subjected to a longitudinal strain. To focus on the first cubic and quartic nonlinear terms of a general potential, we propose an atomic system based on an $\alpha$-$\beta$ Fermi-Pasta-Ulam nearest neighbor interaction.

We perform classical molecular dynamics simulations to investigate the contribution of longitudinal, transversal and flexural modes to the thermal conductance as a function of the $\alpha$-$\beta$ parameters and the applied strain.  We compare the cases where atoms are allowed to vibrate only {\it in} plane (2D) with the case of vibrations {\it in} and {\it out} of plane (3D). We find that the dependence of conductance on $\alpha$ and $\beta$ relies on a crossover phenomenon between linear/nonlinear delocalized/localized flexural and transversal modes, driven by an on/off switch of the strain.

\end{abstract}

\section{Introduction}

The understanding of heat conduction in lattices of interacting particles is still a challenging and fundamental problem of statistical physics \cite{lepri,dhar}. In particular, heat transport in atomic chains has received much attention largely stimulated by its relevance to the nanotechnological applications \cite{nanoph}. Many works on this topic have strongly evidenced that harmonic as well as anharmonic chains can display the so-called anomalous heat conduction, i.e. the heat flow does not follow Fourier's law. In other words, the size dependence of the heat conductivity is one of the most striking feature of anomalous conductors. In this scenario the Fermi-Pasta-Ulam (FPU)\cite{FPU} model and its variants provides an ideal test-bed of the impact of first nonlinear terms of short range general potentials. These models address fundamental issues in statistical mechanics such as the validity of macroscopic laws in low dimensional systems \cite{das1,das2}.

Much of the research on energy transport at the nanoscale has been focused on electronic properties unlike purely thermal (phononic) properties, due to the challenging task of thermal measurements. Low dimensional devices based on Si, B-Ni or C are prominent candidates for nanoscale thermal engineering \cite{baladin,colombo,clivia}, because they exhibit rich thermal physics and potentially offer a route to tailored thermal properties \cite{baladin1,jain}.  Through the confinement of these materials in planar superlattices with one or few atomic planes, many interesting phononic effects emerge. Examples are the role that the flexural phonon band, the effect of 2D superlattices structures, defects or disorder have on the thermal conductivity $\kappa$ \cite{wagner,nika,gosh}. Furthermore, these nanomembranes can operate under different conditions as free-standing or on a substrate, and may also be subjected to different types of mechanical actions that can affect their thermal capability \cite{review}. For example, recent works show that tensile strains applied on Si-nanomembranes, C-based systems as graphene can be used to tune the flexural phonon bands affecting the thermal conductance when the system is small or its asymptotic behavior with the length  \cite{stsz,zhu, pereira, lindsay,xu,nuestroPIP}.

Phononic thermal properties are determined essentially by the vibrations of their constituent atoms. Several fundamental theoretical works devoted to study thermal transport in bidimensional arrays of atoms are based on models that consider mainly in-plane vibrations, i.e. transverse and longitudinal phonon bands \cite{dhar, gendel}. However, recent predictions from numerical works and experiments with nanomembranes show evidence that flexural modes make a significant contribution to $\kappa$ in suspended few-layer graphene and h-BN \cite{review}. That is, vibrations out of plane or flexural phonons play a key role on thermal conduction \cite{xu} and cannot be neglected in the models used for theoretical analysis. In addition, the presence of anharmonic terms in the interatomic potentials plays also a relevant role on the reduction of thermal conductivity due to the change on the distribution of phonon lifetimes through phonon-phonon interactions \cite{zhou}.

In this context, an interesting question is how the interplay between anharmonicities and strain affects the phonon modes and consequently the thermal transport. Our work try to put some light on these phenomena by  a simple model, first allowing atoms to vibrate only in plane (2D case), and secondly we consider the full motion including the out of plane vibrations (3D case). This methodological approach provides an insight into the role of different kind of modes on the thermal transport.

The outline of this paper is the following: in Sec. II we present the model of a nanoribbon with interactions given by the $\alpha$-$\beta$ Fermi-Pasta-Ulam (FPU) potential. In Sec. III we show numerical results for the heat conduction and frequency spectra, analyzing separately the roles of $\beta$, $\alpha$, and the width and length of the nanoribbon. We relate these results with an expansion of the potential when strain is applied. We conclude in Sec. IV by summing up our results and giving some insight into how nonlinearity, phonon bands and strain can be related to the heat transport.

\section{The model}

\begin{figure}[ht]
 	\begin{center}
		\includegraphics[width=0.8\columnwidth]{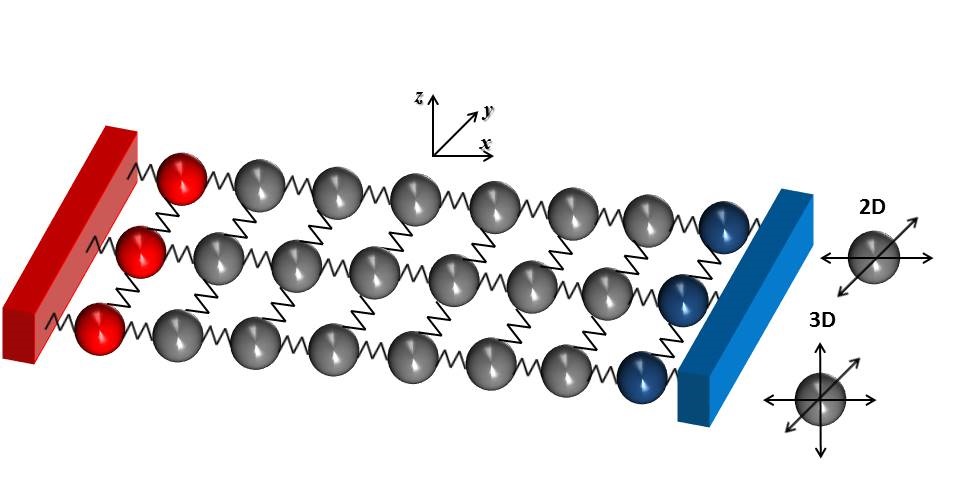}
 	\end{center}  
\caption{Schematic of the system. Each $x$-end is fixed and particles in the first and last columns are coupled to Langevin thermal baths at temperatures $T_L$ (red) and $T_R$ (blue), respectively. Particles can vibrate in $x$-$y$ directions (2D) or in $x$-$y$-$z$ directions (3D).
}
\label{fig:sketch}
\end{figure}

To analyze the impact of nonlinearity on the thermal transport when strain is applied to a suspended membrane, we propose a model that consists of a $N_{x} \times N_{y}$ bidimensional array of identical interacting atoms that can move in the three $x$, $y$ and $z$ directions, vibrating in and out of plane (see Fig \ref{fig:sketch}). 

Particles are located at positions $\textbf{R}_\textbf{i}$ with $\textbf{i}=(i_x, i_y)$ whose equilibrium positions are $\textbf{R}_{0 \bf{i}} = (i_x a_x, i_y a_y, 0)$ with $a_x = a_y = l_0$ the natural equilibrium distance between atoms when no strain is applied. We characterize the motion of the particles by the displacement with respect to their equilibrium position $\textbf{r}_{\bf{i}} = \textbf{R}_{\bf{i}}-\textbf{R}_{0 \bf{i}}$.

The particles in the $x$-ends with $i_{x} = 0$ and $i_{x} = N_{x}+1$ are fixed for all times, while particles in columns $i_x = 1$ and $i_x = N_x$ are in contact with two Langevin thermal reservoirs at different temperatures, respectively. We remark that particles with $i_{y} = 0$ and $i_{y} = N_{y}-1$ are free to move.

The length of the layer is $(N_x + 1) l_{0}$, however when an uniaxial tension is applied along the longitudinal direction, the tension can be parametrized by the change in the lattice constant $a_x > l_{0}$ or by the strain $\epsilon = (a_x-l_0)/l_0$. On the other hand, as the layer is free in the $y$-direction,  $a_y = l_0$ in all cases.

We consider a nearest neighbor interaction up to fourth-order given by an $\alpha$-$\beta$ Fermi-Pasta-Ulam potential that depends on the relative distance $r_\textbf{i,j} = |  \mathbf{R}_{\textbf{i}} - \mathbf{R}_{\textbf{j}} | $. 

\begin{equation} \label{PotFPU}
v(r_{\bf{i,j}}) = \frac{1}{2} k (r_{\bf{i,j}} - l_0)^2 + \frac{1}{3} \alpha (r_{\bf{i,j}} - l_0)^3 + \frac{1}{4} \beta (r_{\bf{i,j}} - l_0)^4  
\end{equation} 
Therefore, the total potential energy $V$ can be written as the sum of pair potentials (bonds) that depend on the atom positions

\begin{equation}
\begin{split}
V= \sum_{\text{bonds}} v(r_{\bf{i,j}}) \\
=\sum_{i_{x}=0}^{N_{x}}\sum_{i_{y}=1}^{N_{y}} v(\left|\textbf{R}_{(i_{x}+1,i_{y})}  - \textbf{R}_{(i_{x},i_{y})}\right|) + \\
\sum_{i_{x}=1}^{N_{x}}\sum_{i_{y}=1}^{N_{y}-1} v(\left|\textbf{R}_{(i_{x},i_{y}+1)} - \textbf{R}_{(i_{x},i_{y})}\right|)%
\end{split}
\end{equation}
being the first term a double sum over longitudinal bonds ($x$ direction), and the second term a double sum over transversal bonds ($y$ direction).

The dynamical evolution of each particle is given by a Langevin type equation of motion with a viscous term proportional to velocity and a decorrelated random force acting on the particles in contact with the reservoirs. The equation of motion for each particle is

\begin{equation} \label{Newton}
\frac{\text{d}^2 \bold{r}_{\bold{i}} }{\text{d} t^2}  = - \frac{\partial V }{\partial \bold{r}_{\bold{i}} } - \gamma_{\bold{i}} \frac{\text{d} \bold{r}_{\bold{i}} }{\text{d} t} + \bold{\xi}_{\bold{i}} (t)
\end{equation}
where $\gamma_{\mathbf{i}} = 1$ for $i_x =1$ and $i_x = N_x$, and zero otherwise. The space-time correlations of the random forces are

\begin{equation}
\langle \xi_{{\bf i},\mu}(t)\,\xi_{{\bf j},\nu}(t') \rangle \:=\: 2\, \gamma\, k_{B}\, T_{{\bf i}}\, \delta_{{\bf i},{\bf j}}\, \delta_{\mu,\nu}\, \delta(t-t')\,.
\end{equation}
where the index $\mu$ and $\nu$ run over $x$, $y$ and $z$ directions. $T_{{\bf i}}$ = $T_{L}$ or $T_{R}$ if $i_{x}=1$ or $i_{x}=N_{x}$, respectively.

For a given realization of the random forces, equations of motion are integrated from a thermal initial state. After a transient time the system achieves a stationary regime where its statistical behavior is constant and thermal properties are calculated. 

Due to Newton's third law and the purely pairwise nature of the forces, the steady state energy current per bond $J_{\bf i,j}$ can be expressed as

\begin{equation}
J_{\bf i,j} = \left\langle \bf{F_{i,j}.(v_{i}+v_{j})} \right\rangle
\label{J1}
\end{equation}
with ${\bf F_{i,j}}=-\nabla_{\bf r_{i}}  v( r_{\bf{i, j}} ) $ the force on particle ${\bf i}$ due to the neighboring particle ${\bf j}$, and brackets indicate time average. Due to the temperature difference in the $x$ direction, the heat current through transversal bonds goes statistically to zero.

In stationary regime, the currents through consecutive longitudinal bonds are statistically equal. The currents for different rows are not necessarily equal due to the asymmetry given by the free boundary conditions. However, this border effect is washed out as $N_y$ is much bigger than one.

Thus, we calculate the mean heat current through the nanoribbon as the average $J$ per row in a transversal plane or column of atoms of the layer as

\begin{equation}
J = \frac{1}{N_{y} (N_{x}-1)}\sum J_{\bf i,j}
\end{equation}
The thermal conductance per row can be calculated from the current $J$ in the horizontal direction determined by temperature gradient $\Delta T / N_x$ between the thermal reservoirs. 

\begin{equation}
C = \frac{J}{\Delta T}
\label{k}
\end{equation}

\section{Results}

In a previous work \cite{nuestroJSM} it was shown that for a general case of particles moving in three directions and when an uniaxial tension is applied, the nonlinear $\alpha$-$\beta$ FPU potential can be expanded around the new equilibrium positions for small displacements. 

If ${\bf i}$ and ${\bf j}$ are the indices of two neighboring atoms in the $x$ direction, the relative longitudinal and perpendicular displacements are defined as $\Delta_{\text{long}} = {\bf r}_{\bf{i,j}}.\hat{x} - a_x$ and $\Delta^2_{\bot} = ({\bf r}_{\bf{i,j}}.\hat{y})^2  + ({\bf r}_{\bf{i,j}}.\hat{z})^2 $ respectively. 


When both displacements are much smaller than the lattice constant $a_x$, we can expand it up to fourth order obtaining

\begin{equation} \label{PotFPUExpanded}
\begin{split}
v(r_{\bf{i,j}}) = v_0 + F_0 \Delta_{\text{long}} + \frac{1}{2} k_{\text{eff}} \Delta^2_{\text{long}} + \frac{1}{2} k_{\perp} \Delta^2_{\perp} + \frac{1}{3} \alpha_{\text{eff}} \Delta^3_{\text{long}} + c_3  \Delta_{\text{long}} \Delta^2_{\perp} + \\
\frac{1}{4} \beta  \Delta^4_{\text{long}} +  \frac{1}{4} \beta_{\perp} \Delta^4_{\perp}  + c_4 \Delta^2_{\text{long}} \Delta^2_{\perp} \ ,
\end{split}
\end{equation}
with

\begin{eqnarray} \label{EffConstants}
v_0 &=& v(a_x) =  \frac{1}{2} k l_0^2 \epsilon^2 + \frac{1}{3} \alpha  l_0^3 \epsilon^3 + \frac{1}{4} \beta  l_0^4 \epsilon^4 \nonumber \\
F_0 &=&  k l_0  \epsilon + \alpha l_0^2 \epsilon^2 + \beta l_0^3 \epsilon^3 \nonumber \\ 
k_{\text{eff}} &=& k + 2 \alpha l_0 \epsilon + 3 \beta l_0^2 \epsilon^2 \nonumber \\ 
\alpha_{\text{eff}} &=& \alpha + 3 \beta l_0 \epsilon \nonumber \\
k_{\perp} &=& \frac{F_0}{l_0 (1 + \epsilon)} \nonumber \\ 
\beta_{\perp} &=& \frac{c_3}{l_0 (1 + \epsilon)} \nonumber \\  
c_3 &=& \frac{1}{2} \frac{k}{l_0 (1 + \epsilon)^2} + \frac{1}{2} \alpha \left( 1 - \frac{1}{(1 + \epsilon)^2} \right) + \frac{1}{4} \beta l_0 \left( \frac{2}{(1 + \epsilon)^2} - 2 + 4 \epsilon \right) \nonumber \\ 
c_4 &=&  - \frac{1}{2} \frac{k}{ l_0^2 (1 + \epsilon)^3} + \frac{1}{2} \frac{ \alpha }{ l_0 (1 + \epsilon)^3} + \frac{1}{2} \beta  \left( 1 - \frac{1}{(1 + \epsilon)^3} \right)
\end{eqnarray}
$F_0$ gives the stress as a function of longitudinal strain $\epsilon$. $k_{\text{eff}}$ is an effective force constant in the longitudinal direction that can increase or decrease with strain, depending on $\alpha$ and $\beta$. $\alpha_{\text{eff}}$ is a term proportional to the cube of the displacement in the longitudinal direction. $k_{\bot}$ is the constant corresponding to the leading term for transversal displacement, which vanishes in case of no strain. In this case the leading term is quartic in the displacement and proportional to $\beta_{\bot}$. $c3$ and $c4$ are two constants which couple the longitudinal and transversal coordinates at third and fourth order in the displacements, respectively. These terms are the responsible for the mixture and coupling of modes.

The equations of motion (Eq.~\ref{Newton}) are integrated with a stochastic Runge-Kutta algorithm. The heat current is calculated after a long enough transient time to guarantee that a steady state is reached.

We consider the equilibrium distance $l_0$ between atoms as unit of length, the mass $m$ of atoms as unit of mass, $\tau_0 = \sqrt{m/k}$ as unit of time, $k l_0^2$ as unit of energy, and therefore $\Theta_0 = k l_0^2/k_{B}$  as unit of temperature.
Although this is a simple theoretical model, we can make some relations with a real carbon-based structure, at least to some orders of magnitude.  For carbon atoms in graphene $l_0 \approx$ 0.14 nm and $k \approx$ 650 nN/nm typically, giving $\Theta_0 \approx 10^6$ K and $1/\tau_0 \approx$ 180 THz for temperature and frequency units, respectively.
We work with $T_{L}=3.8\cdot 10^{-4}$ and $T_{R}=2.0\cdot 10^{-5}$, which would correspond to 380 K and 20 K, respectively, or an average temperature of 200 K.
Expanding the Tersoff-Brenner potential \cite{Tersoff}, usually used for carbon-based or Si-based nanosystems, around the equilibrium position up to fouth order we obtain the dimensionless values $\alpha \approx -5.5$ and $\beta \approx 16.9$.
Even though we present some results for these particular parameters, we address in general the role of $\alpha$ and $\beta$ on the thermal conductance of the strained layer.

\begin{figure}[h]
\begin{minipage}{0.5\textwidth}
\subfloat{\includegraphics[width=6cm,height=5cm]{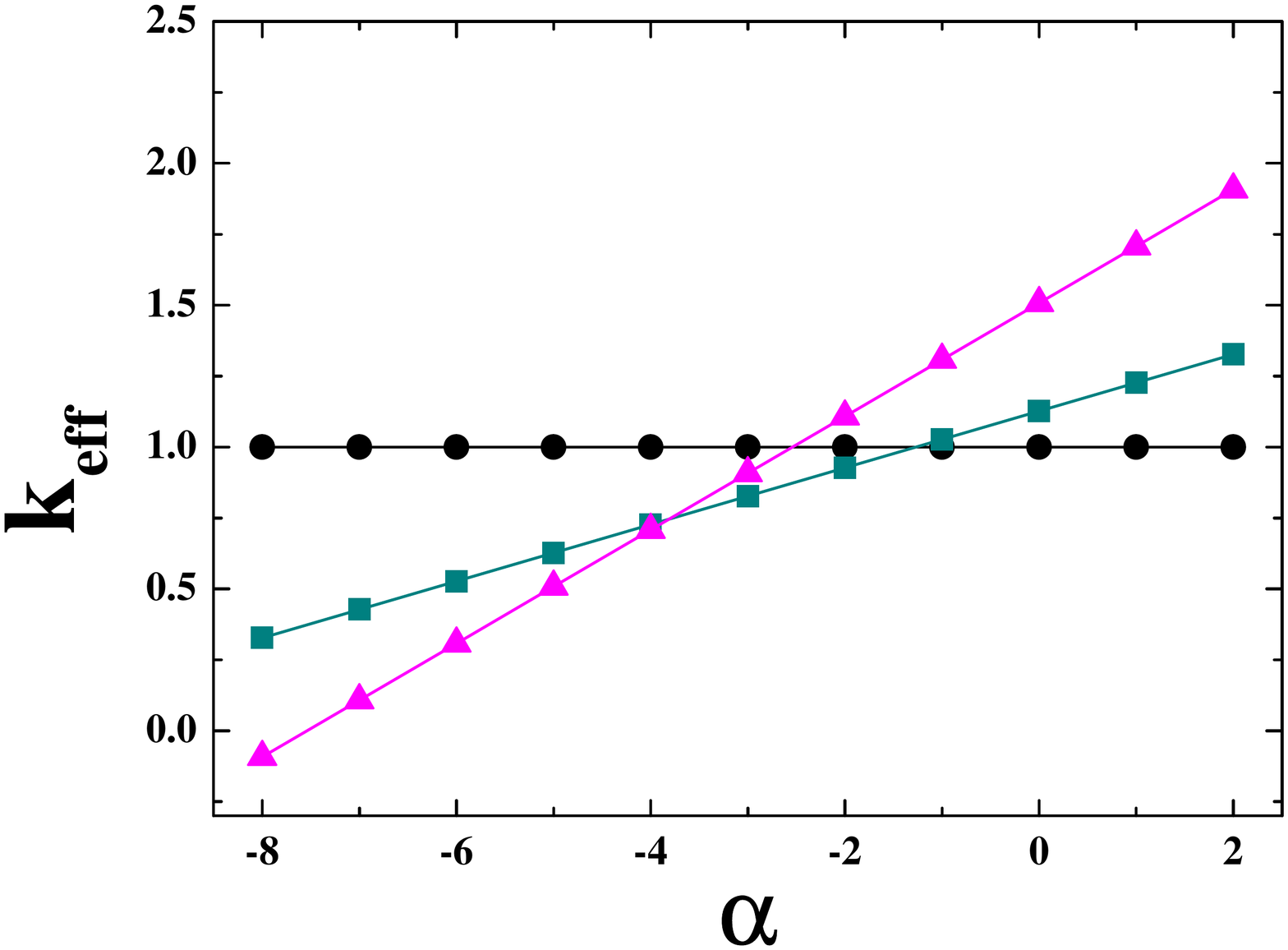}}
\end{minipage}%
\hfill
\begin{minipage}{0.5\textwidth}
\subfloat{\includegraphics[width=6cm,height=5cm]{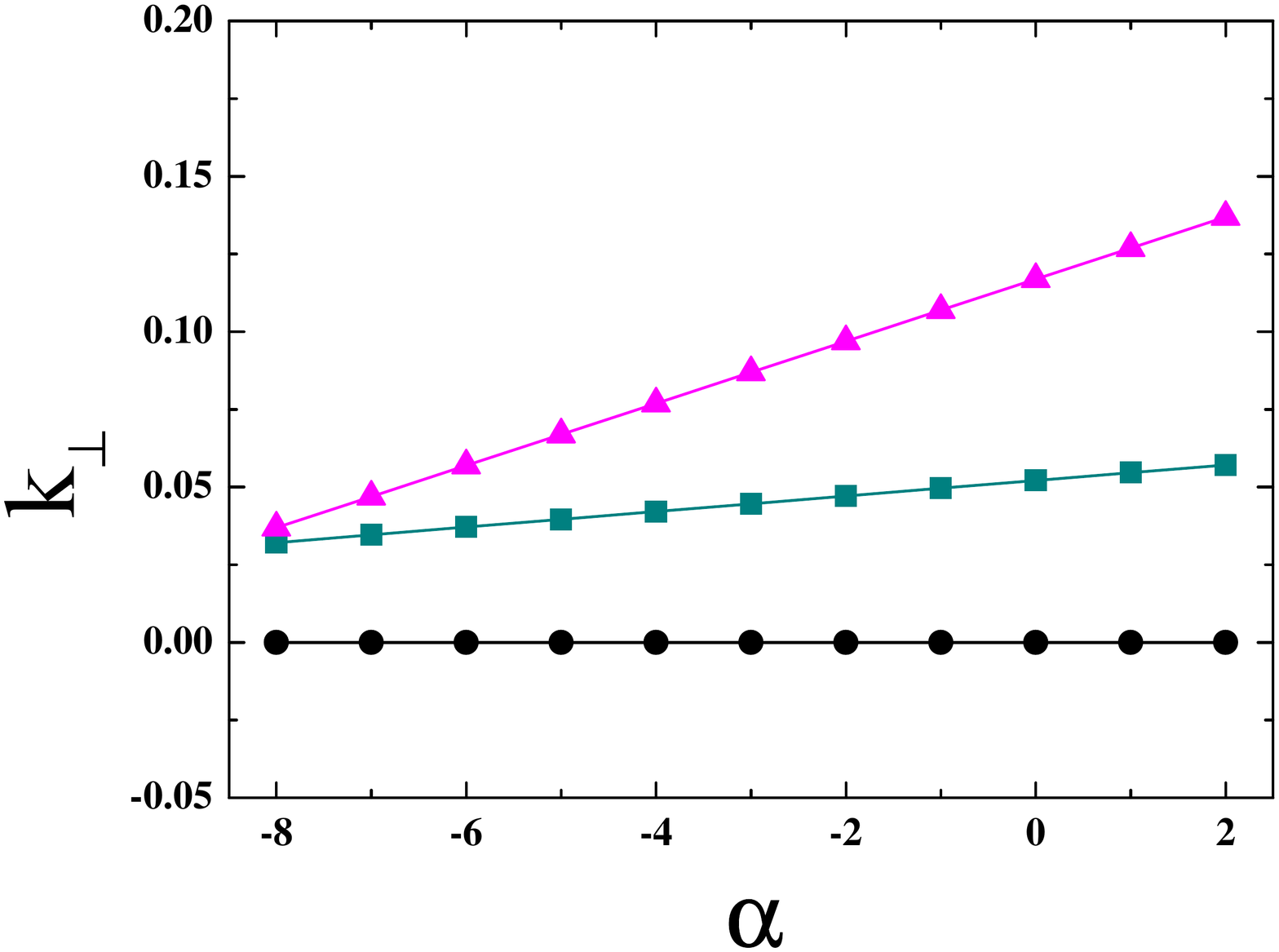}}
\end{minipage}%
\vfill
\begin{minipage}{0.5\textwidth}
\subfloat{\includegraphics[width=6cm,height=5cm]{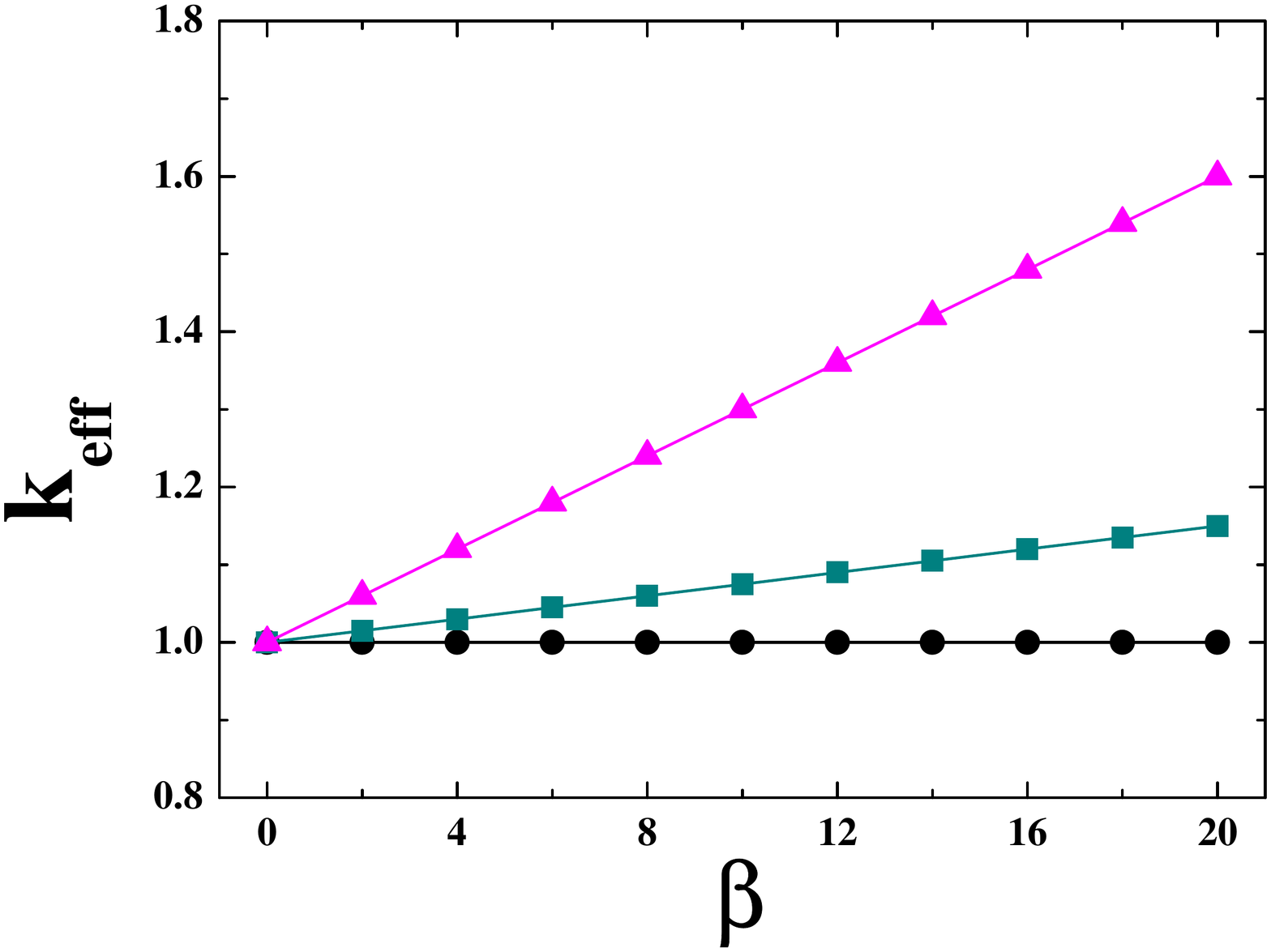}}
\end{minipage}%
\hfill
\begin{minipage}{0.5\textwidth}
\subfloat{\includegraphics[width=6cm,height=5cm]{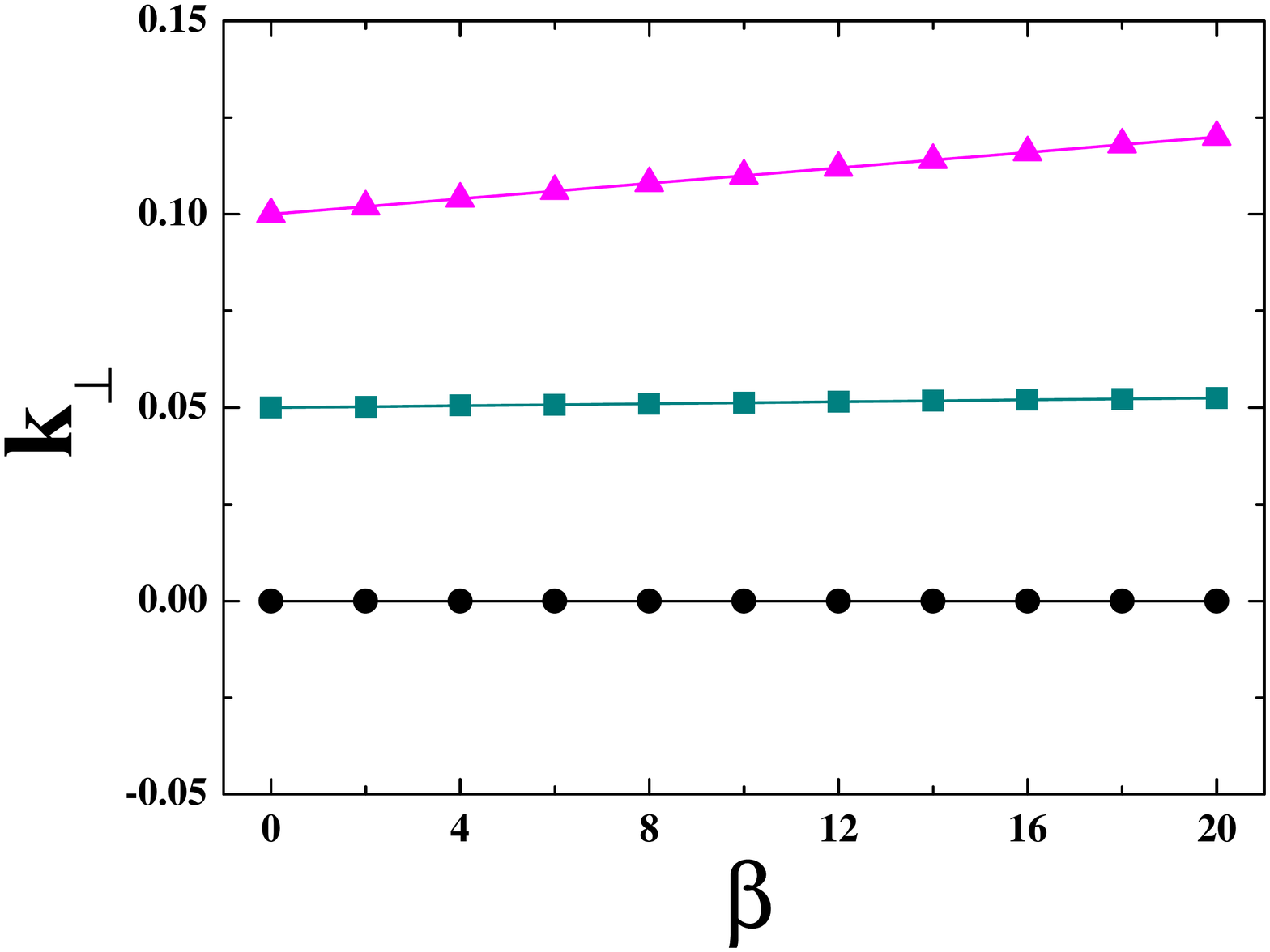}}
\end{minipage}%
\vfill
\caption{Effective longitudinal $k_{\text{eff}}$ and transversal $k_{\perp}$ elastic constants as a function of $\alpha$ (upper row) and $\beta$ (bottom row). Black dots corresponds to $\epsilon=0$, green squares to $\epsilon=0.05$ and magenta triangles to $\epsilon=0.1$.}
\label{fig:keff}
\end{figure}

\subsection*{\textbf{Role of $\beta$}}%

\begin{figure}[ht]
 	\begin{center}
 		\includegraphics[width=0.46\columnwidth]{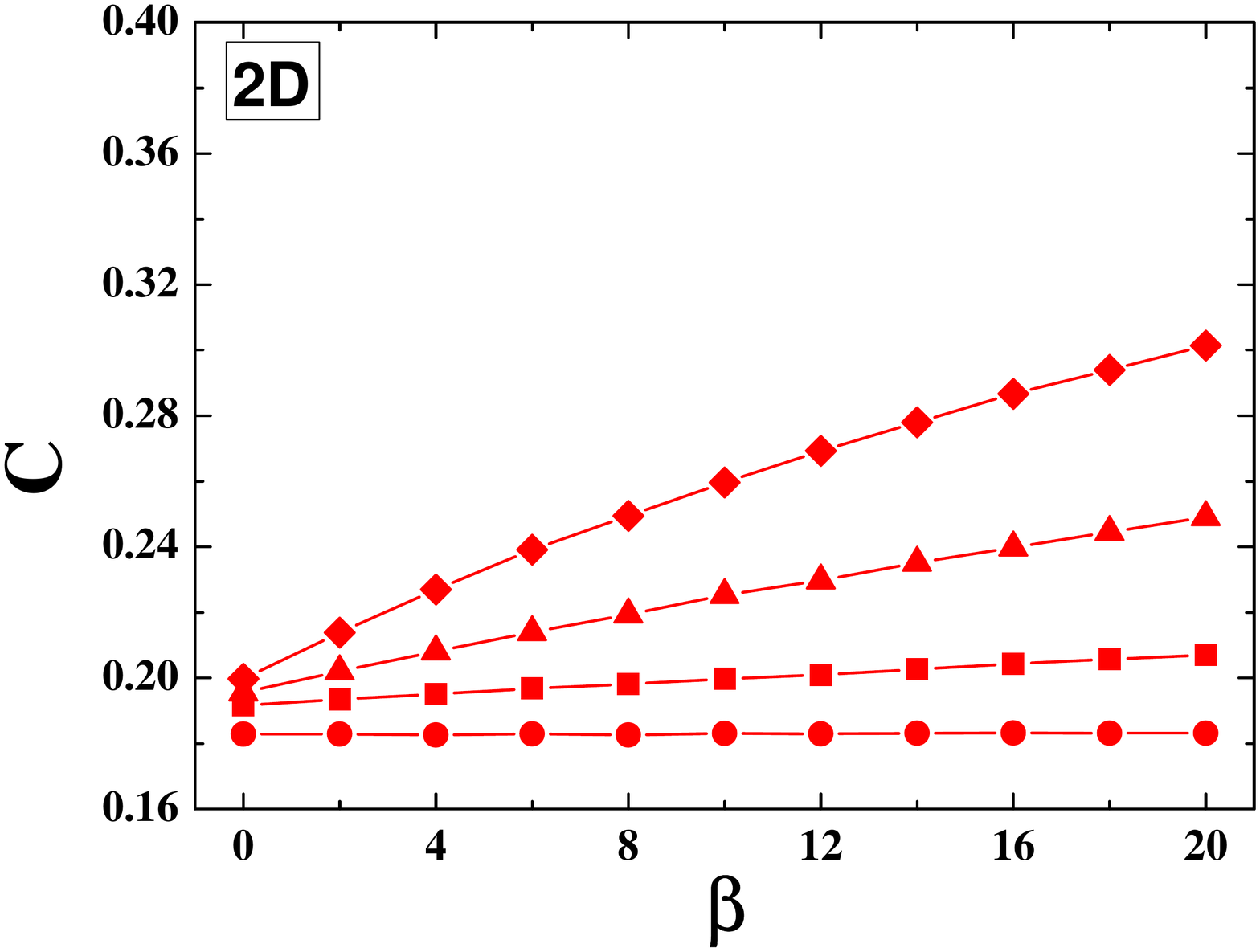}
 		\includegraphics[width=0.46\columnwidth]{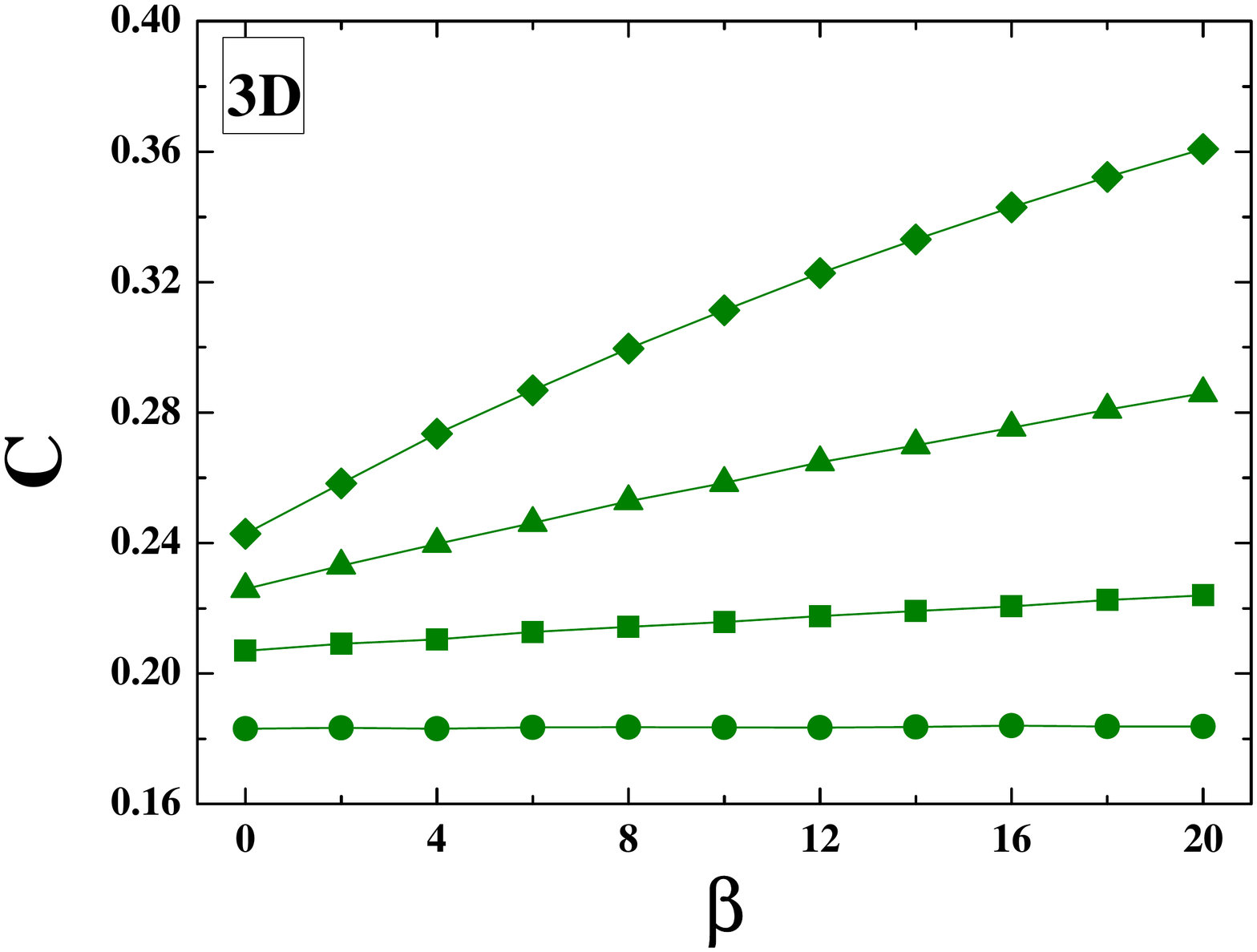}
 	\end{center}  
\caption{Conductance $C$ as a function of $\beta$ for 2D (left) and 3D (right) cases. For both panels different values of $\epsilon$ are considered: $\epsilon=0$ (dots), $\epsilon=0.05$ (squares), $\epsilon=0.1$ (triangles) and  $\epsilon=0.15$ (diamonds). $N_x =20$, $N_y =20$, $T_{L}=3.8\cdot 10^{-4}$, $T_{R}=2.0\cdot 10^{-5}$ and $\alpha=0 $.}
\label{fig:beta}
\end{figure}

For a symmetric potential, the quartic anharmonic term represents the lower non-linear order of an attractive force. It is expected to play a relevant role on the heat conduction since it is related to phonon-phonon scattering processes. However, it is interesting to analyze if the impact of this term is enhanced or not when a strain is applied, and in particular when due to the dimensionality of the system flexural modes are considered. In Fig. \ref{fig:beta} the thermal conductance $C$ for the $2D$ and $3D$ cases is plotted as a function of $\beta$ and for different values of $\epsilon$. To do a more exhaustive analysis of the role of $\beta$ we assume at this point that no cubic term is present ($\alpha=0$).

We find that for a non strained ribbon ($\epsilon= 0$), $C$ takes almost the same value both in $2D$ and $3D$. From Eqs.~\ref{EffConstants} we observe that the effective constants depend on $\beta$ only when strain is applied. If $\epsilon$ is close to zero constants $c_3$ and $c_4$ mostly depend on $k$ and so the coupling between longitudinal ($x$) and transversal ($y$-$z$) displacements.

\begin{figure}[H]
\begin{minipage}{0.5\textwidth}
\subfloat{\includegraphics[width=6cm,height=5cm]{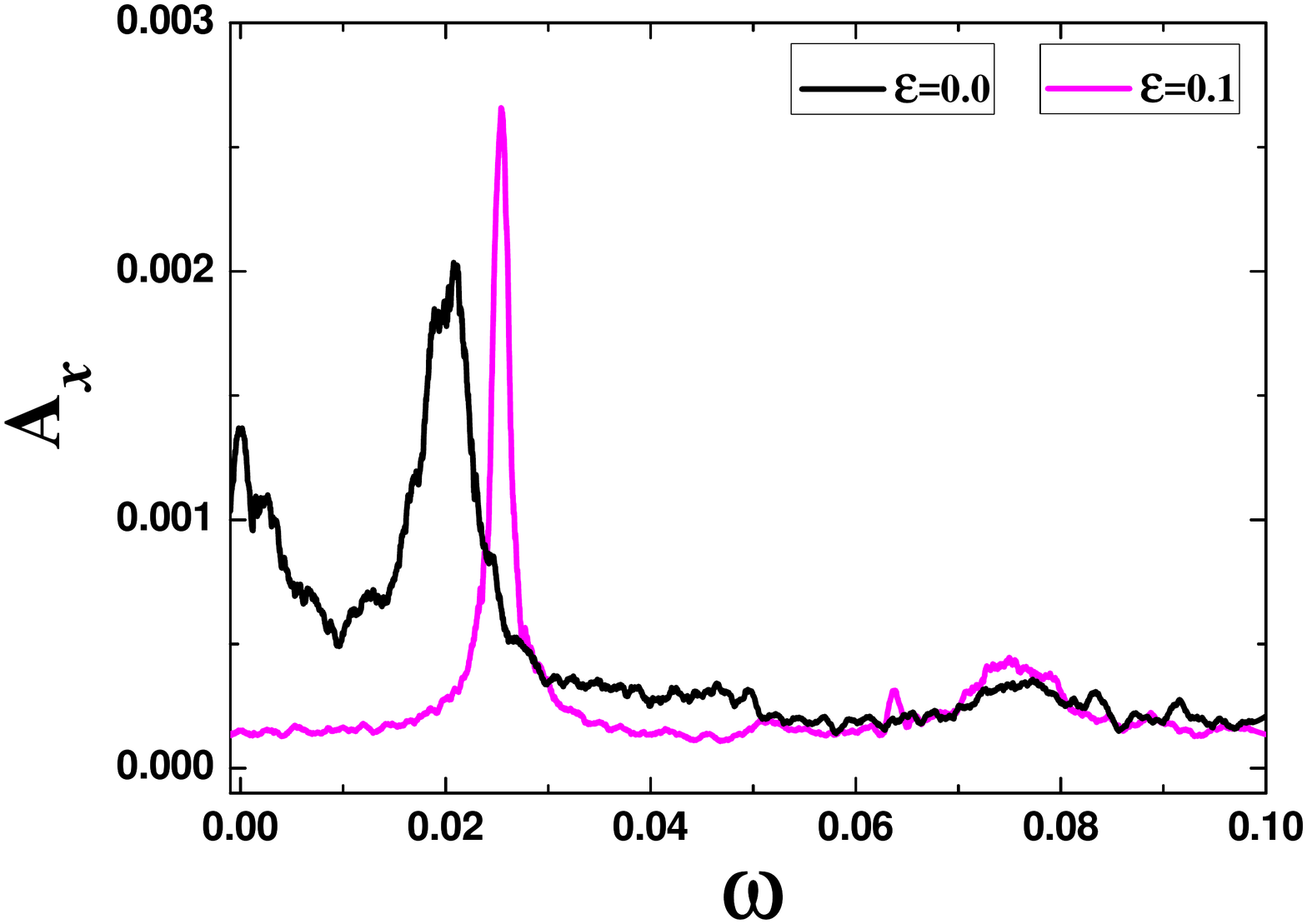}}
\end{minipage}%
\hfill
\begin{minipage}{0.5\textwidth}
\subfloat{\includegraphics[width=6cm,height=5cm]{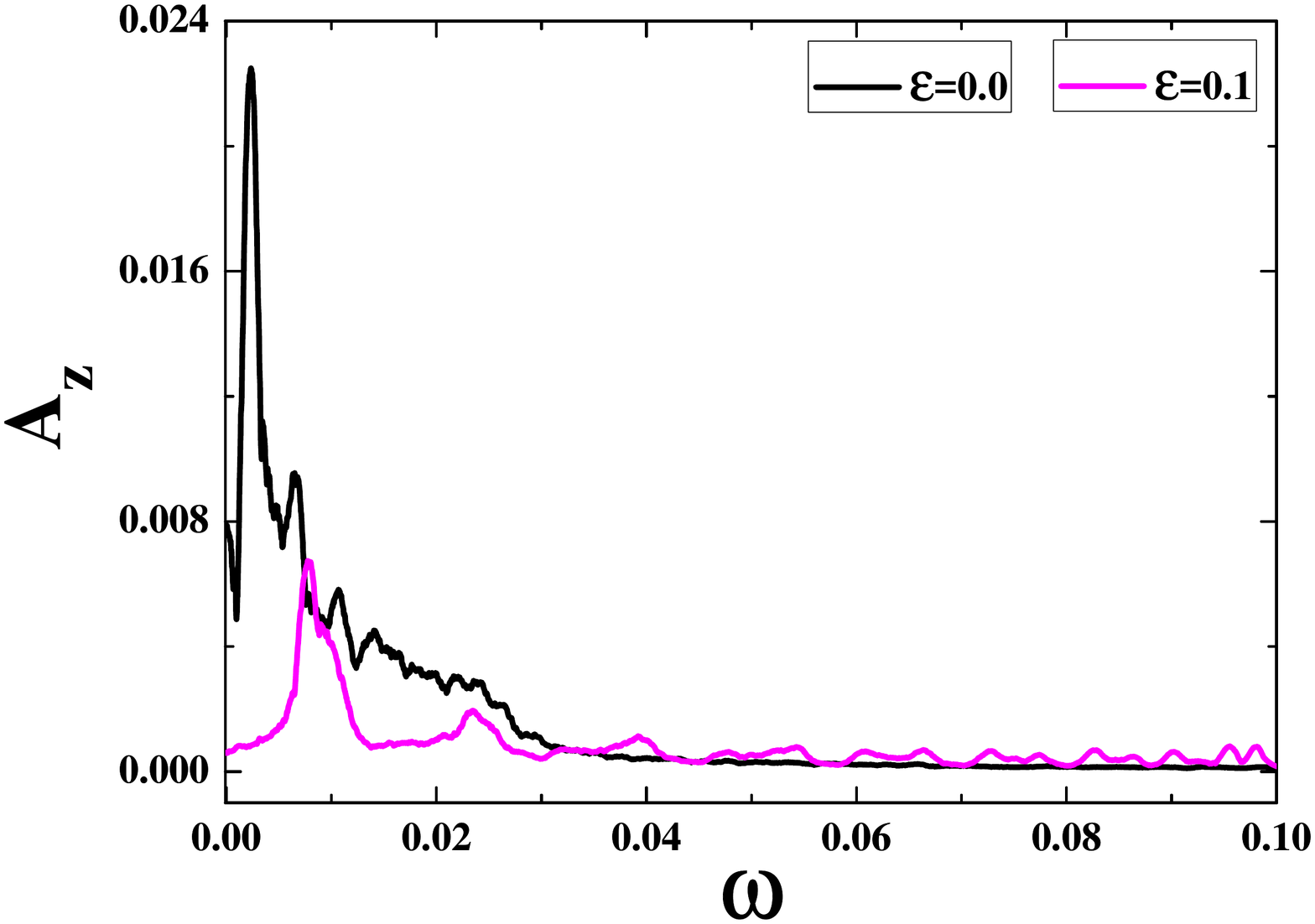}}
\end{minipage}%
\vfill
\begin{minipage}{0.5\textwidth}
\subfloat{\includegraphics[width=6cm,height=5cm]{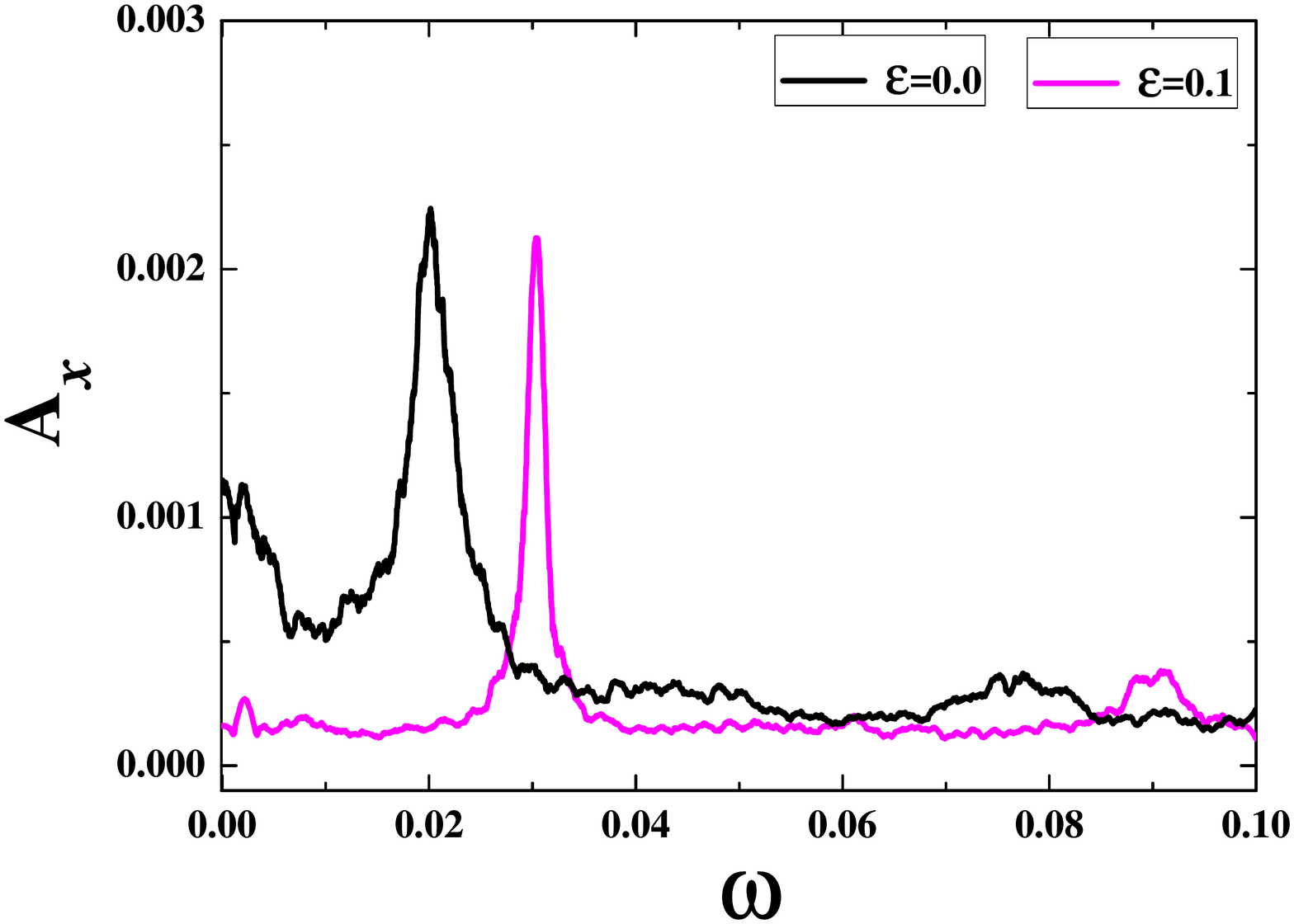}}
\end{minipage}%
\hfill
\begin{minipage}{0.5\textwidth}
\subfloat{\includegraphics[width=6cm,height=5cm]{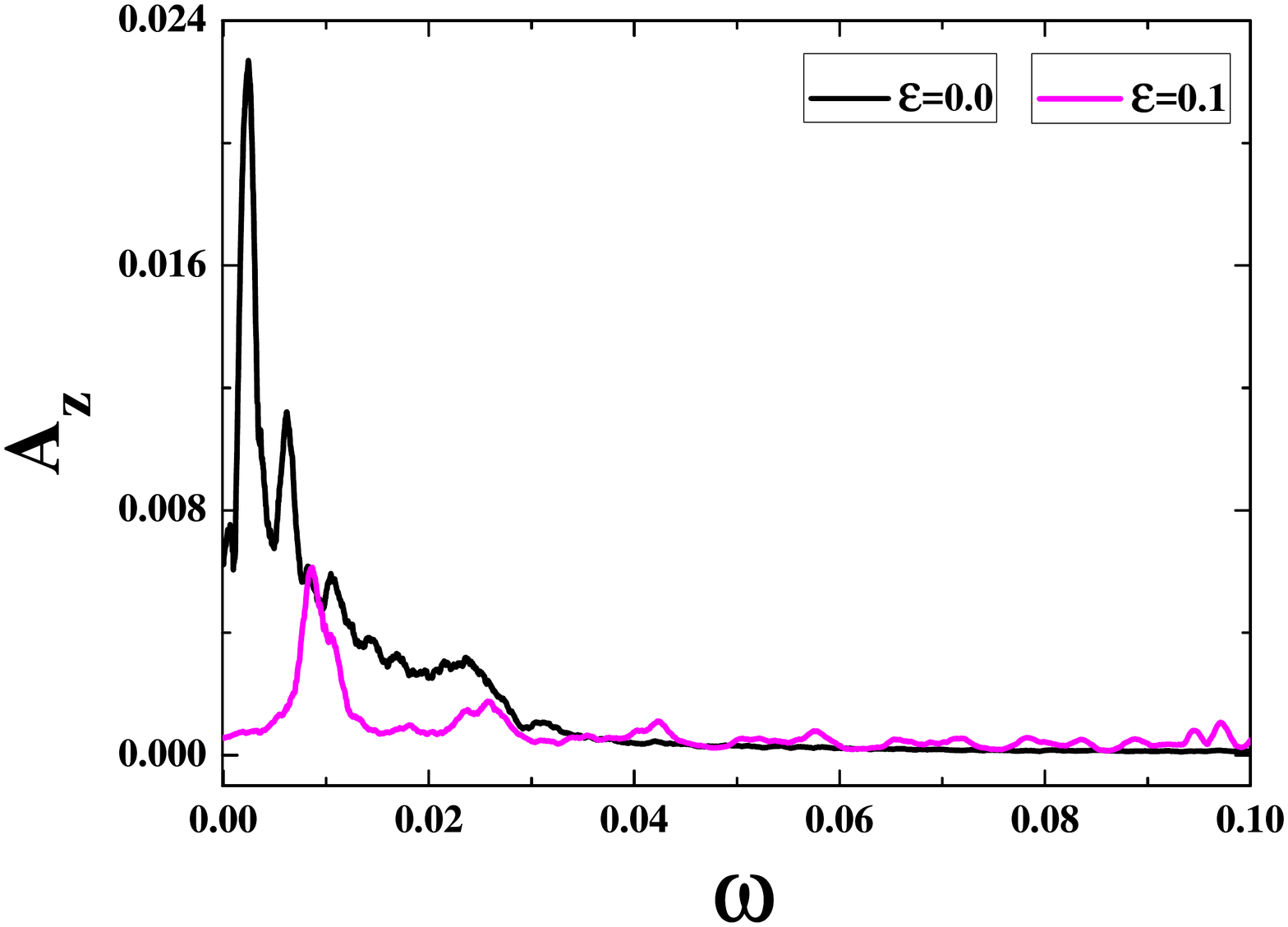}}
\end{minipage}%
\vfill
\caption{Spectra of longitudinal (left pannels) and flexural (right panels) displacements for the 3D case. Top row corresponds to $\beta=4$ and bottom row to $\beta=20$. In each panel, black solid line corresponds to non strained case ($\epsilon=0$) and magenta solid line corresponds to $\epsilon=0.1$. In all cases $\alpha = 0$,  $N_x =20$, $N_y =20$, $T_{L}=3.8\cdot 10^{-4}$, $T_{R}=2.0\cdot 10^{-5}$ }
\label{fig:spectraB}
\end{figure}

To analyze the contribution to conductance from different types of modes, we plot in Fig. \ref{fig:spectraB} the spectra of the longitudinal and flexural displacements of a central atom for two values of $\beta$. We observe that for the unstressed case, flexural modes are present with a considerably energetic contribution and at a frequency that is almost independent of $\beta$. As the transversal effective elastic constant is zero, the flexural modes are essentially non linear. Moreover, from Fig. \ref{fig:beta} it seems that their contribution to thermal conductance is almost negligible, suggesting that these flexural modes are localized.

On the other hand when $\epsilon$ is finite, the effective constant $k_{\perp}$ is also finite and increases with $\beta$. Thus the system behaves closer to a harmonic one and the main contribution comes from the linear modes.  In addition as $c_3$ and $c_4$ also increase, the coupling between in/out of plane modes is stronger. Therefore, the energy stored in the flexural modes, now essentially linear and non localized, contributes to thermal conductance increasing $C$ with $\beta$. These effects are more evident for larger strains because the effective coupling constants are also larger (see Eq.~\ref{PotFPUExpanded}). Consequently the thermal conductance in 3D is enhanced respect to 2D case, as it is shown in Fig. \ref{fig:beta}.

Regarding the spectra, the mean peak of the longitudinal band is shifted towards larger values of the frequency for finite $\epsilon$, corresponding to modes with higher energy. This shift is greater as long as $\beta$ increases. In addition, the second peak of the flexural spectra and the main peak of the longitudinal spectra are located in the same frequency band. This is an evidence of the coupling between in plane/out of plane modes that contributes to thermal conductance.

\subsection*{\textbf{Role of $\alpha$}}

Usually in an expansion around the equilibrium distance of a real interatomic two-body potential, the cubic term is negative. This is due to the characteristic core repulsion which makes the potential to rapidly increases at short distances, while it soften at longer distances, where it usually tends to zero. Therefore the $\alpha$ term is important to take into account the asymmetry of interatomic potentials. Here we study the dependence of the conductance on $\alpha$ while keeping $\beta = 16.93$ fixed, in order to always have a bound motion.

We checked that for this value of $\beta$ and the used range of values for $\alpha$, there is only one equilibrium distance between atoms. A second equilibrium distance would produce a different equilibrium conformation of the ribbon.

\begin{figure}[ht]
 	\begin{center}
 		\includegraphics[width=0.46\columnwidth]{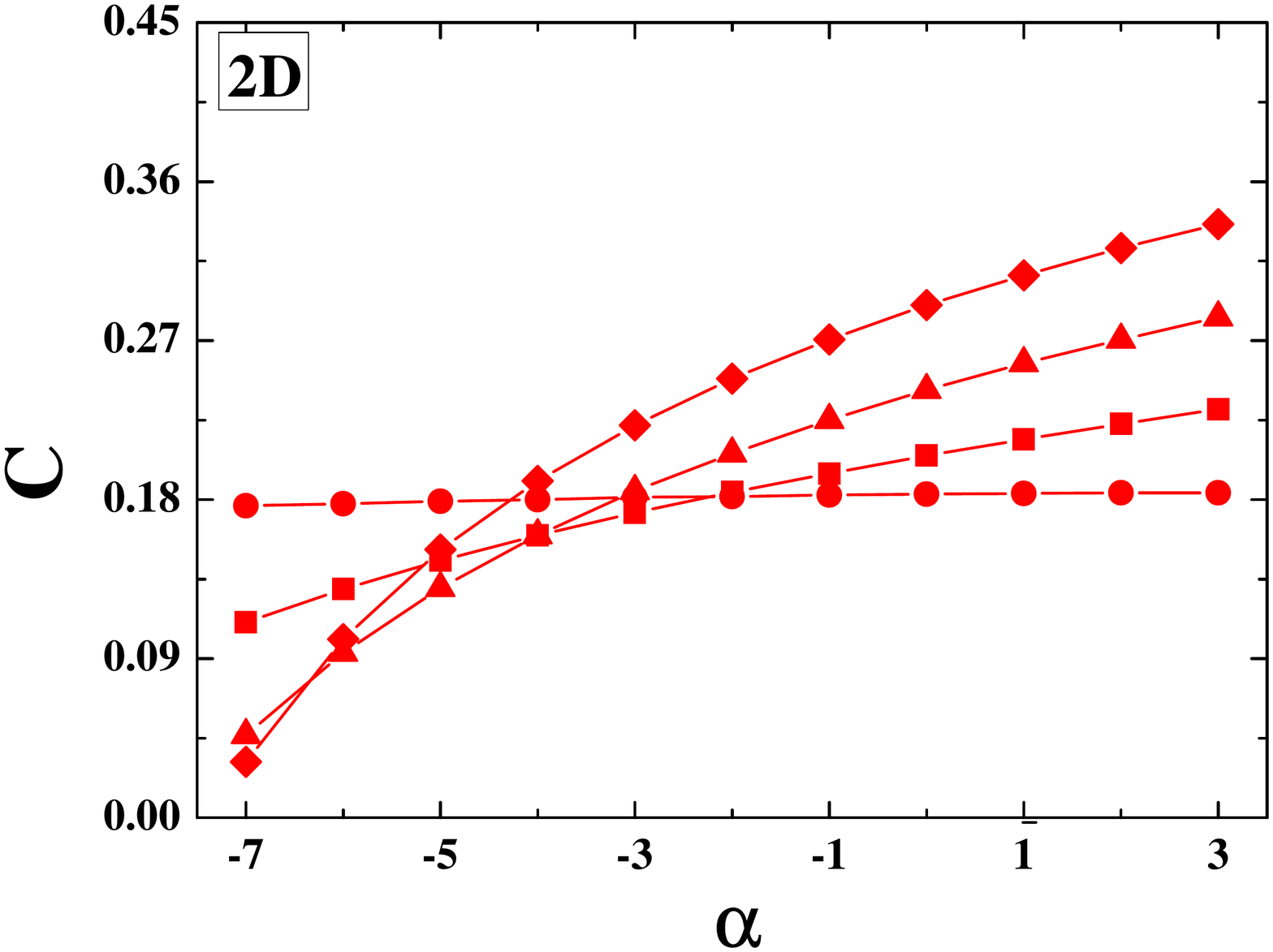}
 		\includegraphics[width=0.46\columnwidth]{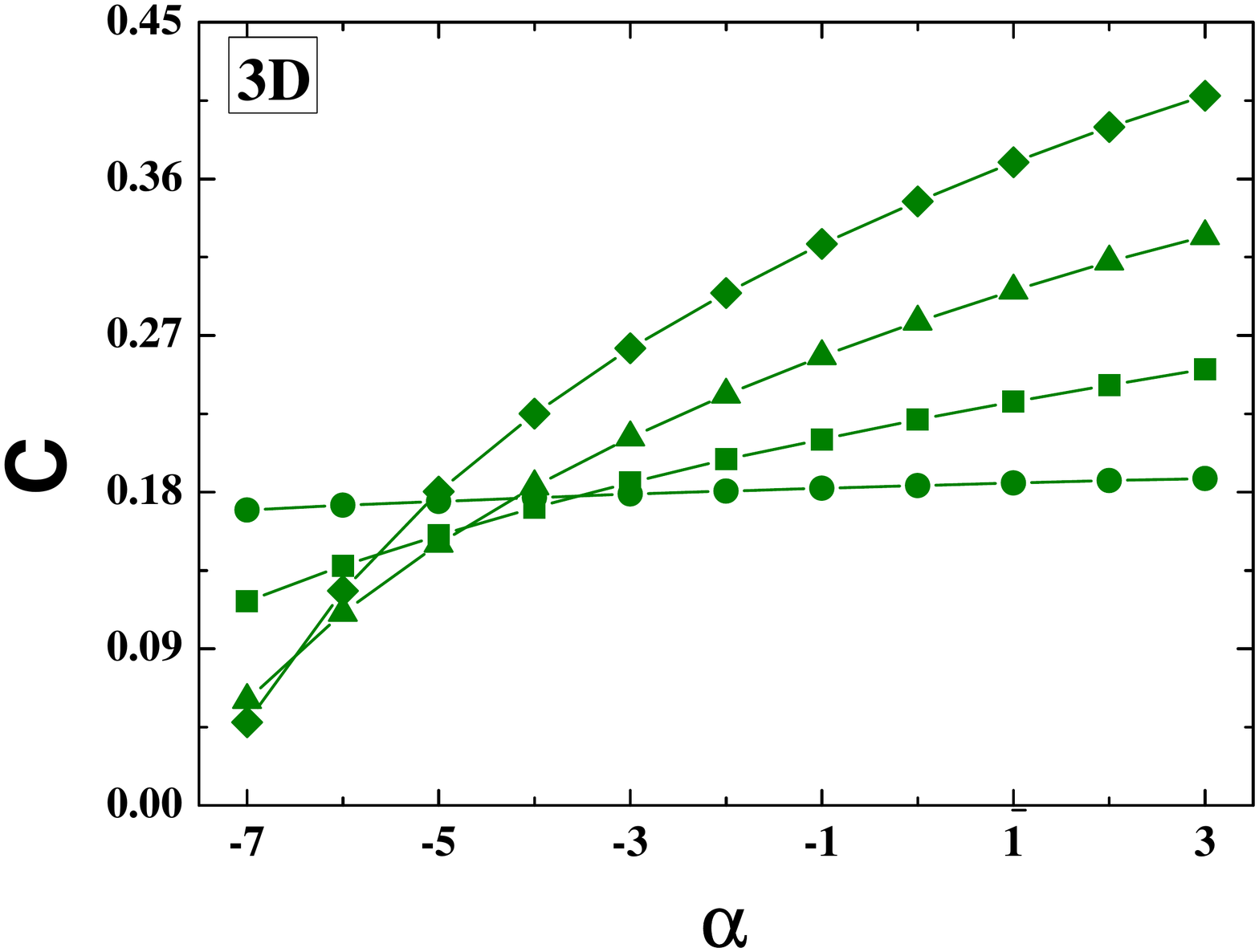}
 	\end{center}  
\caption{Conductance $C$ as a function of $\alpha$ for 2D (left) and 3D (right) cases. For both panels different values of $\epsilon$ are considered: $\epsilon=0$ (dots), $\epsilon=0.05$ (squares), $\epsilon=0.1$ (triangles) and  $\epsilon=0.15$ (diamonds). $N_x =20$, $N_y =20$, $T_{L}=3.8\cdot 10^{-4}$, $T_{R}=2.0\cdot 10^{-5}$ and $\beta=16.93$.}
\label{fig:alfa}
\end{figure}

Without strain on the ribbon, the conductance does not depend on $\alpha$ (see Fig.\ref{fig:alfa}). Moreover, the same value $C_0 \approx 0.18$ is observed for 2D and 3D motion, pointing to a negligeable contribution of the flexural modes to the conductance, although they can store vibrational energy.

On the other hand, as soon as the ribbon has some strain, the conductance displays a monotonic increase with $\alpha$. For a given strain, there exist a critical value $\alpha^*$ for which $C$ is bigger than $C_0$ for $\alpha > \alpha^*$, while $C < C_0$ for $\alpha < \alpha^*$. This critical value $\alpha^*$ is more negative as long as the strain increases. This effect is more evident in the 3D case than in the 2D case. For example in the 3D case and for $\epsilon = 0.15$, $\alpha^* \approx -5 $ and in the 2D case $\alpha^* \approx -4 $.

These behaviors can be understood from Eqs. \ref{EffConstants}, specially from the value of $k_{\text{eff}}$. In Fig. \ref{fig:keff} we see the dependence of this effective harmonic constant as a function of $\alpha$ for different strains, and it strongly correlates with the behavior of the conductance. When $k_{\text{eff}}$ becomes smaller than one, the ribbon softens in the longitudinal direction.

The constant $k_{\perp} \approx \epsilon \ k $ for small strains, and it contributes to both the increasing frequencies of transversal and flexural modes (in $y$ and $z$ directions respectively). So, this can also explains the larger conductance in the 3D case compared to the 2D.

\begin{figure}[h]
\begin{minipage}{0.5\textwidth}
\subfloat{\includegraphics[width=6cm,height=5cm]{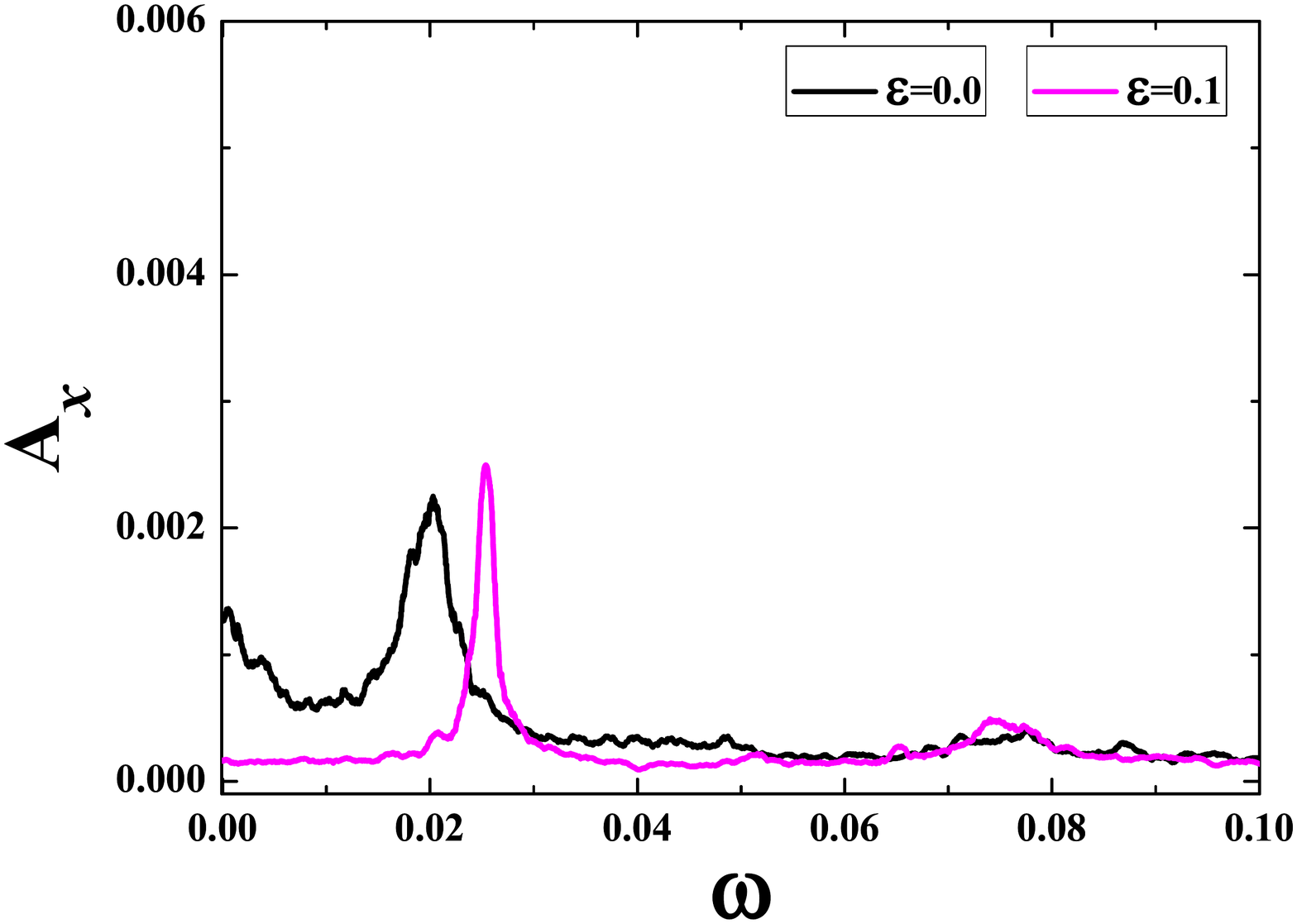}}
\end{minipage}
\hfill
\begin{minipage}{0.5\textwidth}
\subfloat{\includegraphics[width=6cm,height=5cm]{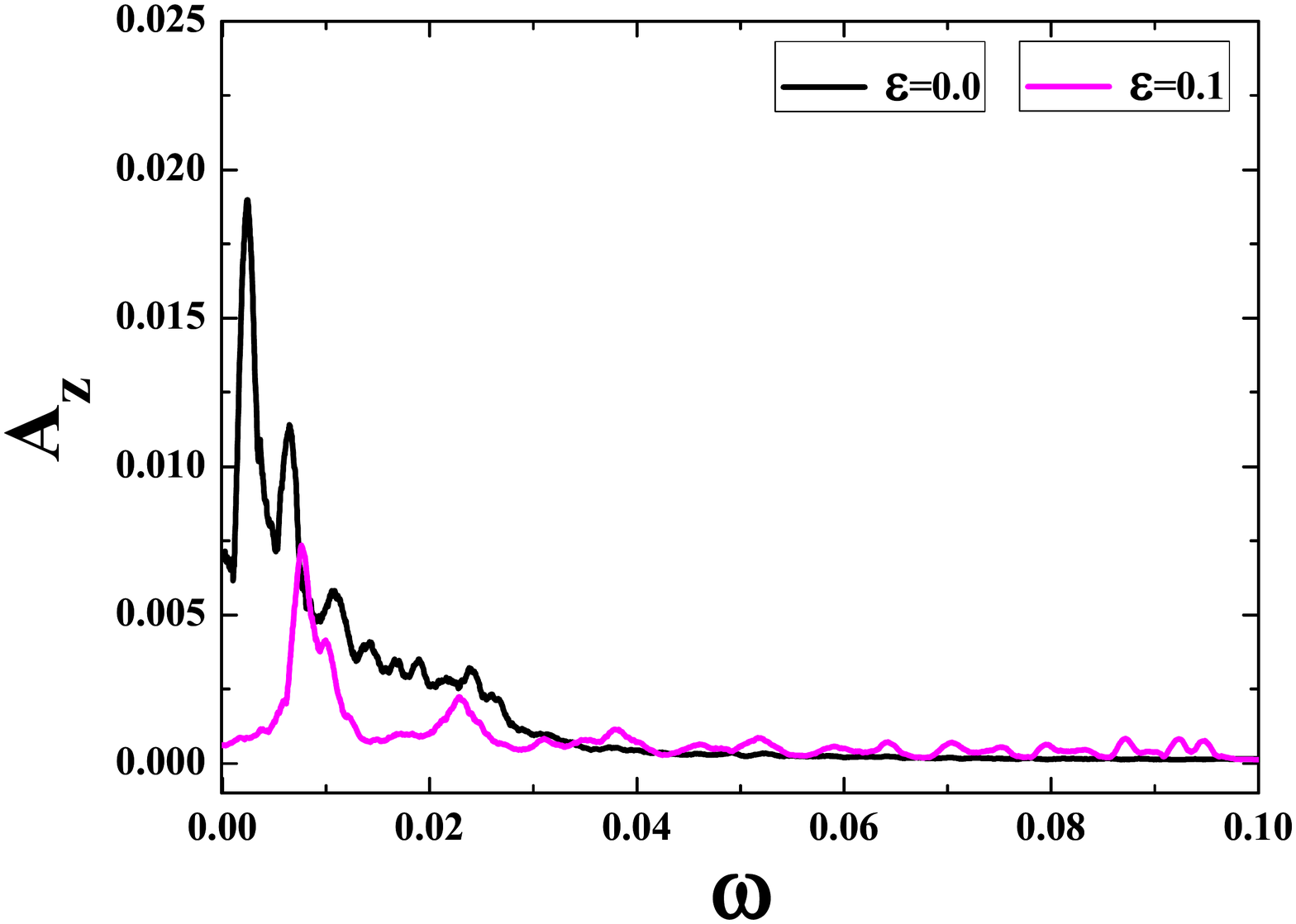}}
\end{minipage}%
\vfill
\begin{minipage}{0.5\textwidth}
\subfloat{\includegraphics[width=6cm,height=5cm]{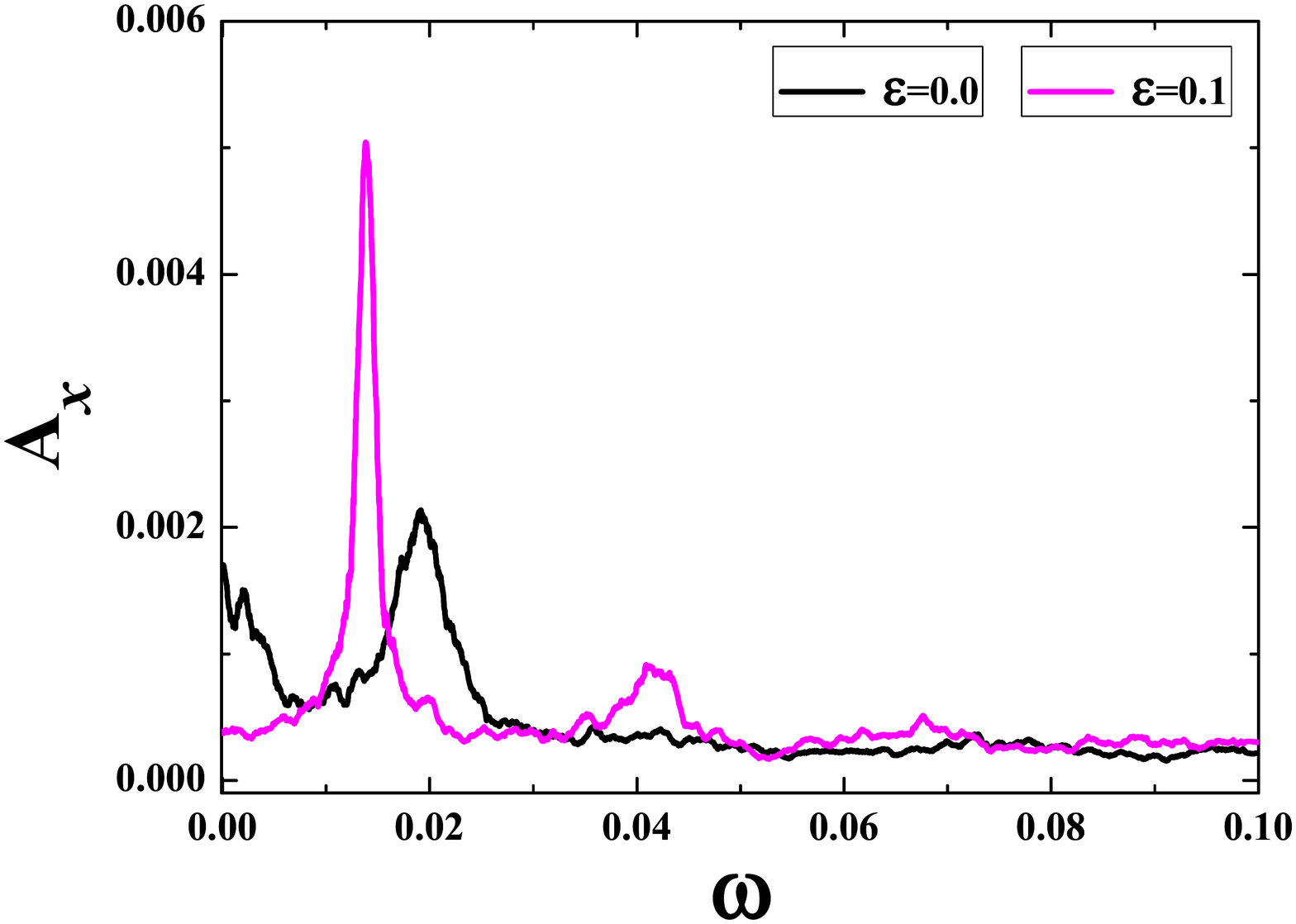}}
\end{minipage}%
\hfill
\begin{minipage}{0.5\textwidth}
\subfloat{\includegraphics[width=6cm,height=5cm]{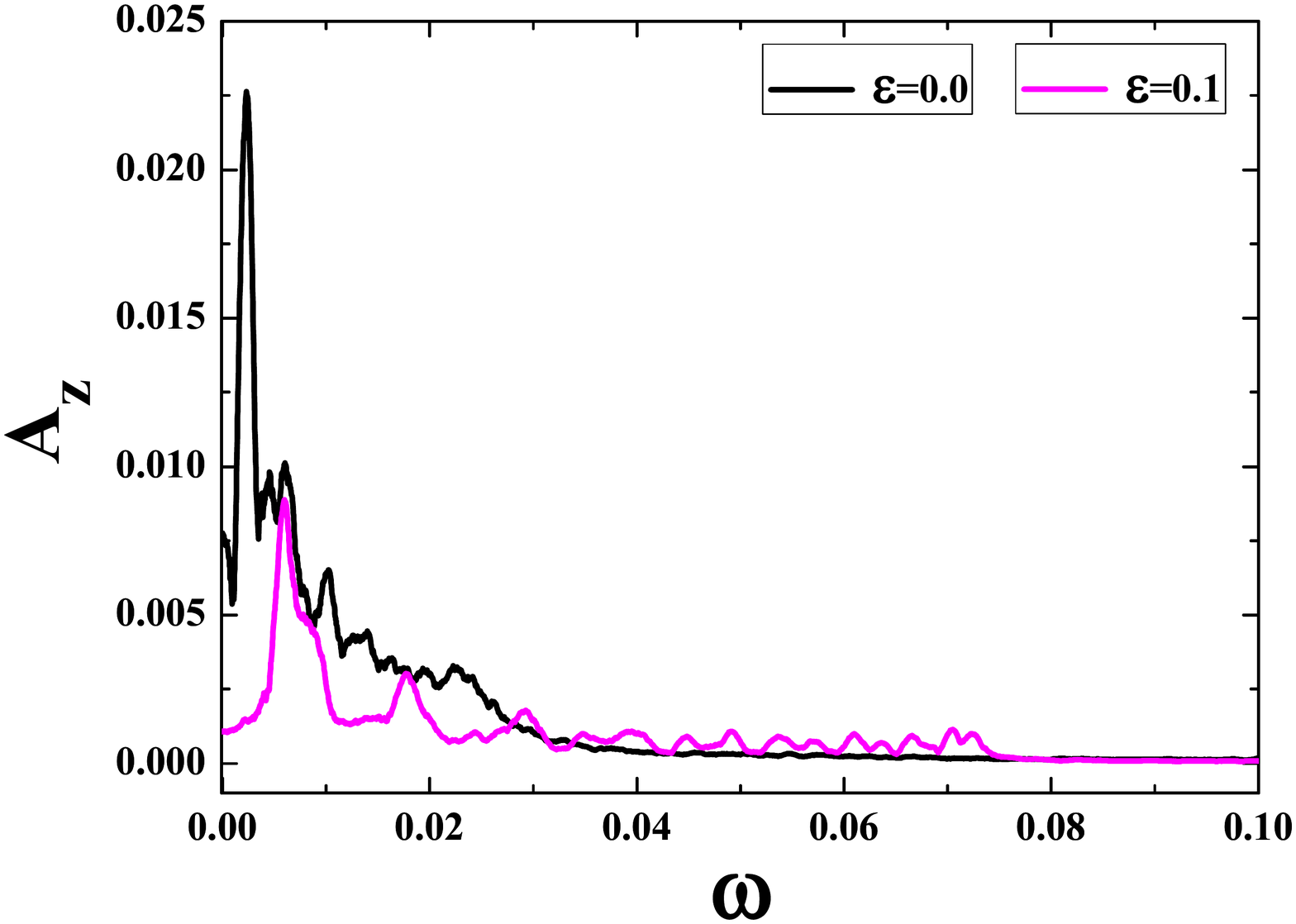}}
\end{minipage}%
\vfill
\caption{Spectra of longitudinal (left pannels) and flexural (right pannels) displacements for the 3D case. Top row corresponds to $\alpha =-2.0$ and bottom row to $\alpha = -6.0$. In each panel, black solid line corresponds to non strained case ($\epsilon = 0$) and magenta solid line to $\epsilon = 0.1$. In all cases $\beta = 16.93$,  $N_x =20$, $N_y =20$, $T_{L}=3.8\cdot 10^{-4}$, $T_{R}=2.0\cdot 10^{-5}$.}
\label{fig:spectraA}
\end{figure}

In Fig. \ref{fig:spectraA} we plot the spectra of the  $x$ and $z$ displacements for one atom in the center of a ribbon with $N_x \times N_y$ = 20 $\times$ 20. We consider the 3D case with and without strain.

For $\epsilon = 0$, only minor differences are observed among the spectra for different values of $\alpha$, which is compatible with the already observed constant conductance. For $\epsilon = 0.1$ we observe a smaller frequency of the first peak for $\alpha = -6$ as compared to $\alpha = -2$. This correlates again with the values of $k_{\text{eff}} = 0.969$ and $k_{\perp} = 0.088$, for $\alpha = -2$, and $k_{\text{eff}} = 0.569$ and $k_{\perp} = 0.052$, for $\alpha = -6$.

\subsection*{\textbf{Conductance vs $N_x$ and $N_y$}}

To investigate the dependence of the conductance with the size of the ribbon, we fix the interatomic potential to $k = 1, \alpha = -5.45, \beta = 16.93$. These values are given by the expansion of the Tersoff-Brenner potential for carbon atoms, as it was explained in the model.

\begin{figure}[ht]
 	\begin{center}
 		\includegraphics[width=0.46\columnwidth]{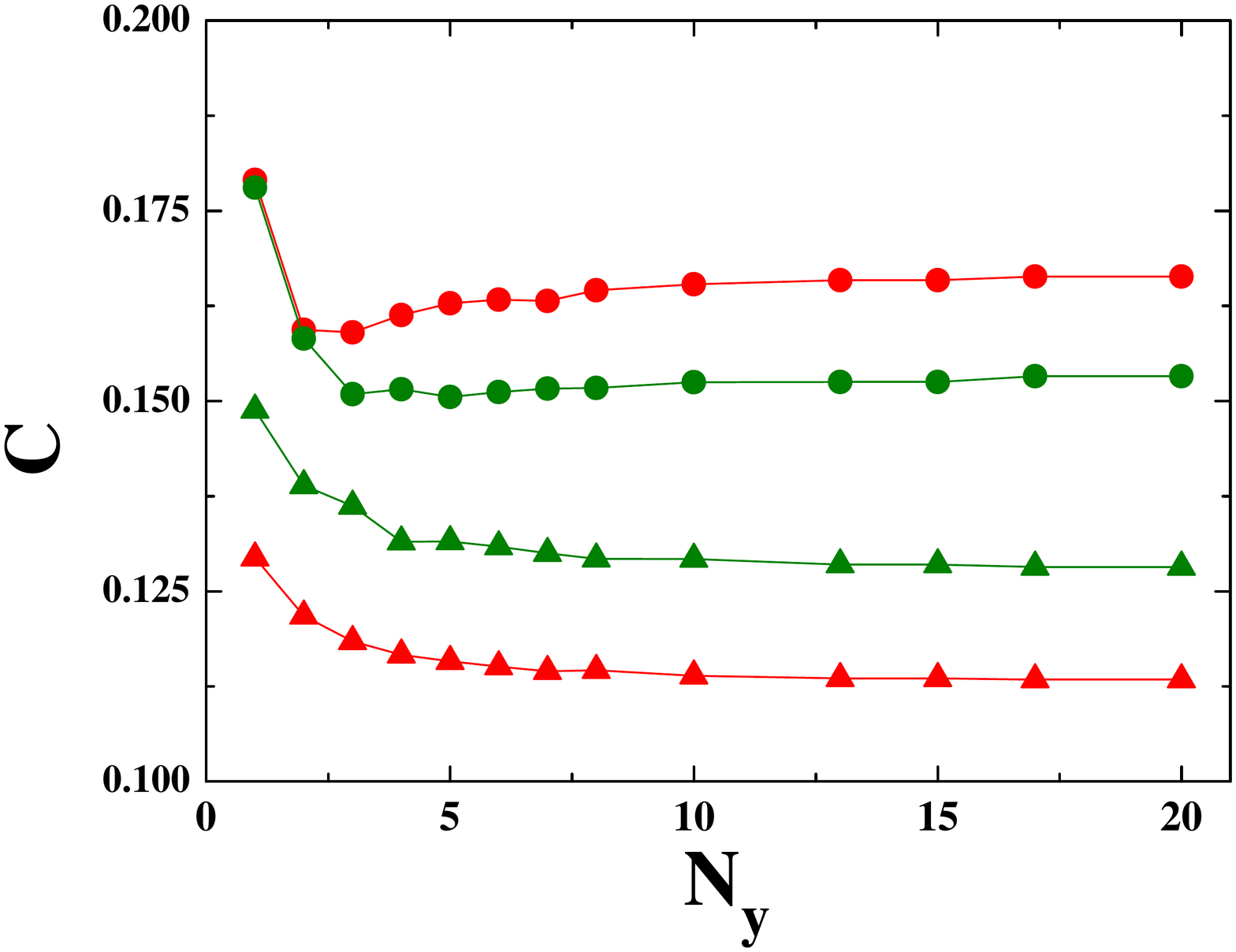}
 		\includegraphics[width=0.46\columnwidth]{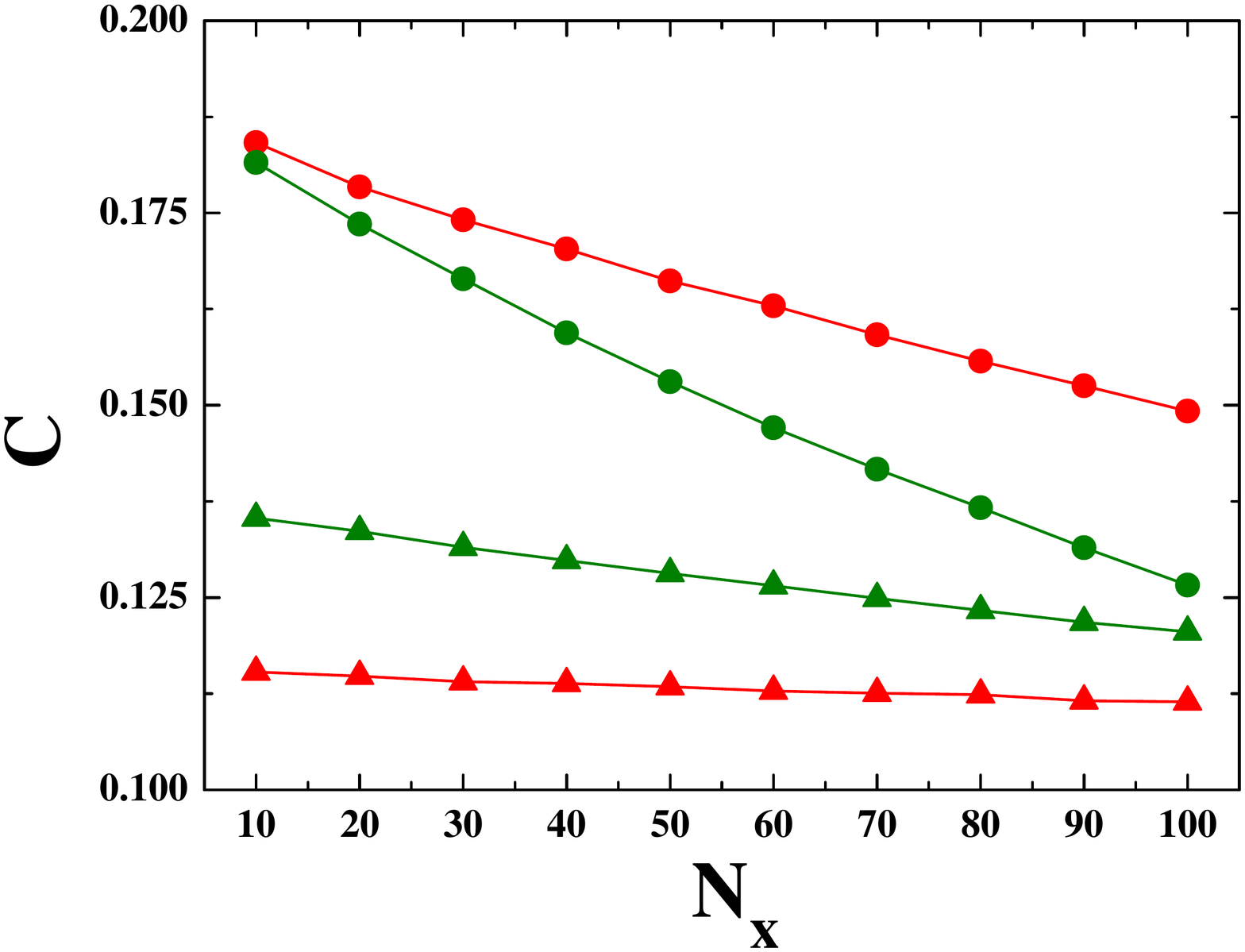}
 	\end{center}  
\caption{Left panel: $C$ as a function of the width $N_y$ for 2D (red) and 3D (green) with $N_x =50$. Right panel: $C$ as a function of the length $N_x$ for 2D (red) and 3D (green) with $N_y =20$.  In both figures, circles correspond to $\epsilon=0$ and triangles to $\epsilon=0.1$. In all cases $\alpha=-5.45$, $\beta = 16.93$, $T_{L}=3.8 \cdot 10^{-4}$, $T_{R}=2.0\cdot 10^{-5}$.}
\label{fig:size}
\end{figure}

We first fix the length of the ribbon to $N_x = 50$, and we vary its width $N_y$. The results are in the left plot of Fig. \ref{fig:size}. We observe a decay of conductance for small values of $N_y$, which is faster for the non strained configuration $\epsilon = 0$. The case $N_y = 1$ corresponds to a single wire of atoms, being qualitatively different from a ribbon. In all cases the conductance (per row) converges to a constant value for $N_y \approx 10$. This suggests that some eventual effects on the conductance coming from the free lateral boundaries of the ribbon are washed out rapidly as $N_y$ grows.

Secondly, we fix the width of the ribbon ($N_y = 20$) and we vary its length $N_x$ (see right plot in Fig.\ref{fig:size}). We observe in general that conductance is lower for the strained case as compared to the non strained. This is due to the particular value of $\alpha$ which is lower than $\alpha^*$.

However, there is a qualitative difference: for the strained ribbon the conductance barely decreases with length, while for the non strained case the decreasing is more evident.

We explain this behavior from the presence of harmonic modes in all directions when $\epsilon$ is finite, which are delocalized and entail a constant conductance \cite{crystal}. For the non strained case, some transversal and all flexural modes are non-linear, which are mainly localized. These modes do not contribute to the conductance producing more scattering of phonons, and conductance reduces with the length of the system. Moreover, this can also explain why in the 3D case the conductance is smaller than in the 2D case. All flexural modes are non linear and they reduce the conductance given by the harmonic longitudinal and tranversal modes.

\section{Conclusions}

We have investigated the role of nonlinearity in the interatomic potential on the thermal conductance when a nanoribbon is subjected to a longitudinal strain.

We considered an $\alpha$-$\beta$ Fermi-Pasta-Ulam model as a general expansion of any potential that, besides a harmonic term, includes nonlinear cubic and quartic terms. Unlike other recent theoretical works, we considered a model for the potential that depends on the absolute distance between atoms and where atoms can vibrate in the three directions, as it is expected in a real system. Consequently phonon bands are constituted by longitudinal, transversal and flexural modes.

We studied the role of these modes on the thermal transport from the expansion of the total potential energy around the new equilibrium positions. We analyzed how $\alpha$, $\beta$ and longitudinal strain affect effective coupling constants and consequently the thermal conductance.

We found that in the absence of strain the conductance in 3D is smaller than in 2D, unlike the strained case. This puzzling fact relies on the changeover from nonlinear to linear modes when the strain is 'switched on'. In the non strained case all flexural and some transversal modes are non linear, and presumably most of them are localized. Therefore, they do not contribute to the conductance, or they even reduce it due to scattering processes with the remaining linear modes. In other words from a regime where phonons can be scattered to a regime where they can propagate ballistically.

It is interesting to note that this transition takes place even when the size of the system ($N_x$ and $N_y$) is smaller than the mean free path of the phonons.  This can bring innovative solutions to control the thermal conductance when operating with small size devices by tuning properly the applied strain. Or in other words, a thermal control in nanostructures based on the localization/delocalization of nonlinear/linear phonon modes.

Our proposed atomistic model is a suitable approach to understand thermal transport in nanosytems mediated by acoustic phonons. Many challenging technological applications arise on the use of nanowires and nanotubes as strain sensors by inducing the system into a thermally nonlinear vibrational regime \cite{nuestroJSM, nanosensor}. Our model shows a complementary technological implication. Even at low temperatures and small system sizes where phonon-phonon scattering may not be relevant, the dependence of conductance on the strain allows, among other things, to extract information about characteristic parameters of the first non linear terms in the interatomic potential. 

More theoretical work should be done to better understand how  localization affects the interplay between different linear/nonlinear vibrational modes and their contribution to the thermal conductance in strained systems. However, our results provide some valuable suggestions in this direction.


\section*{Acknowledgments}
This work is supported by PIO CONICET-UNGS grant.


\end{document}